# An Information-oriented Model of Multi-Scale (Feedback) Systems

Ada Diaconescu[*], Louisa Jane Di Felice[**], Patricia Mellodge[***]
[*]Telecom Paris, [**]Autonomous University of Barcelona, [***]Hartford University

## Motivation & Objective

Multi-scale structures are prevalent in both natural and artificial systems, as they can handle increasing complexity [3],[4],[5]. Several terms are employed almost interchangeably across various application domains to refer to the multi-scale concept – e.g., hierarchy, holarchy, multi-level, multi-layer, nested, embedded, micro-macro or coarse graining [6],[7],[8],[9]. While the concrete meanings behind these terms may differ slightly, several core commonalities persist across all cases. In this position paper we aim to highlight these common features of the multi-scale concept, as a preliminary basis for a generic theory of multi-scale systems. We discuss the concepts of *scale* and *multi-scale* systems in general, and then of *multi-scale feedback systems* in particular, focusing on the role played by information in such systems. Our long-term objective is to develop a general theory of multi-scale feedback systems, applicable across all domains dealing with complex systems.

## Previous Work

In previous work, we have highlighted the common features of multi-scale structures across various natural and artificial domains [10], and started formalising these via a generic design pattern – multi-scale abstraction feedbacks (MSAF) [11]. In brief, we model multi-scale systems as ensembles of *information flows*, which are merging and splitting to generate various information *abstractions* (bottom-up, micro-to-macro) and information *reifications* (top-down, macro-to-micro). In [17], we study the impact of inter-scale timing characteristics on the resulting macro-properties and behaviour of multi-scale systems.

## What is Scale?

From across dictionary entries[1,2,3] the term "scale" has two main meanings that are relevant to our study. These definitions are somewhat interrelated:

- Related to **measurement** – either the units of measurement (e.g., metric scale) or the instruments of measurement (e.g., weighting scale).
- Related to a **ratio** between a real object and a model of the object (e.g., a map scale). Here, scale can also be used as a verb – i.e., to scale-up or -down – meaning that something pertaining to the scaled object increases or decreases proportionally with that with respect to which it is scaled. E.g., scaling-up a business, or a computing process, means that it can deal with more incoming requests while using a proportional amount of resources.

We associate scale to the measurement of a system of interest, where such measures are then used to create system models, at a certain ratio. Models may then be used to help scale-up the coordination of system entities.

---

[1] Merriam-Webster dictionary online: https://www.merriam-webster.com/dictionary/scale
[2] Math vocabulary: https://www.splashlearn.com/math-vocabulary/measurements/scale
[3] Wikipedia: https://en.wikipedia.org/wiki/Scale

Based on these considerations, we define **scale** as: *the granularity of observation of a targeted object* (Figure 1). Granularity can represent, e.g., an interval, range, abstraction level, or frequency. Hence, importantly, scale is a property of the observation and not of the observed object. Moreover, scale necessarily implies the existence of an observer, performing the observation. Thus, scale is relative to an observer's perspective during an observation process. As we will see later, the observer does not have to be an external human observer – it can be another external system or even an internal system entity (it is any entity that can take information from the system, and adapt based on that information). This is consistent with the notion of scale in Allen and Starr's Hierarchy Theory [12].

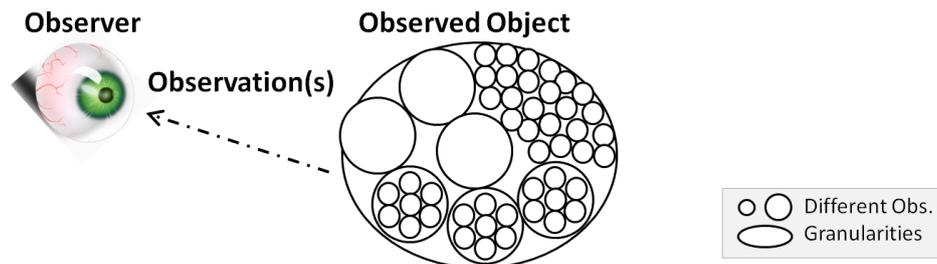

Figure 1: Scale is the granularity of observation of a targeted object by an observer

In this view, an information scale is related to the kind of information acquired by an observer about an observed object. A "higher" scale refers to a larger, or coarser, granularity of observation of a targeted object (more abstraction). A "lower" scale, on the other hand, implies a smaller, or finer, granularity of observation of the same targeted object (more detail). In multi-scale systems in general, higher scales provide more abstract information about larger scopes of the observed object; while lower scales provide more detailed information about narrower scopes of the observed object. This limits the amount of information observed, stored and processed at each scale.

## Information Systems

Observation is the acquisition of **information** about an object of interest. Generally, such observation may or may not lead to the analysis of the acquired information, and to some sort of action, change or adaptation in the observer (the receiver of that information). As we focus on a phenomenological (or behaviourist) view of systems, we are only interested in cases where observation does lead, eventually, to some observable change (even if after some delay, via storage and later processing).

Hence, for the purpose of our study, we consider the definition of information as *an observable change in an object that propagates and triggers adaptation in an observer*. This view is consistent with the semantic definition of information in [1] and with the "information as adaptation" view proposed by [13]. Differently from a purely syntactic definition quantifying information in terms of a flow of bits, the semantic view that we build upon expands on the material properties of information, placing attention on the function of the information flow (i.e., adaptation in the observer). This does not mean that information cannot be quantified in terms of material flows, but identifies those flows as information flows only when they are observed, producing a change. An **information flow**, thus, is a series of events, or changes, that lead to changes in an observer (the recipient of that information flow). We can model our generic system of observed objects, observations and observers (Figure 1) as an ensemble of information flows influencing each other: an observed information flow propagates changes via an observation information flow leading to changes into an observer information flow (Figure 2). We link this informational system model to physical systems as follows (extension of considerations in [10]).

Fundamentally, a physical object of interest is some sort of **process** – i.e., a series of events, or changes in observable quantities, or variables, happening in some order [2]. When a process keeps its identity, tied to a set of variables and their values, unchanged for a relatively long period (from the observer's perspective) we consider the process to be an **entity**; and consider its few variables that do change to define the entity's state (e.g. a stone, keeping its structure static for a long period, while changing only a few of its variables, like position, speed and temperature, in the eye of a person observing it).

Information flows are linked to physical processes, or entities, by the fact that any piece of information must rely upon, or be encoded onto, a physical substrate (even if the same piece of information may be encoded onto various physical substrates). This is consistent with the findings in [14], where a piece of information about a physical object requires a minimum amount of energy (or negative entropy – negentropy) to be extracted. This information can then be employed to create some energy in return (negentropy), while of course abiding to the second law of thermodynamics.

Based on the above considerations, we model **a system as a network of information flows**, which change, observe and adapt to each other repeatedly. Each information flow may be observed by other information flows, via yet other information flows. The system is populated by information flow triplets: i) an observed information flow, ii) an observation information flow and iii) an observer information flow (Figure 2). Information scales may vary across such information flow triplets within a system. To simplify, we only refer in the following to observer-observed pairs, implicitly including the observation flow connecting them. While we keep the three kinds of information flows separate, we can consider the system as a network of information flows that run in parallel, merge into each other, and split at various points (e.g., just like we distinguish between rivers with different names, whereas they merely represent a large network of water flows).

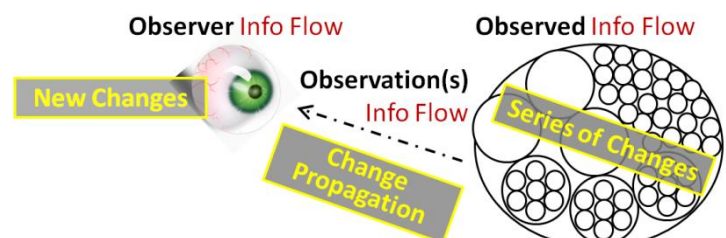

Figure 2: System model focused on information flows: observer, observation and observed object

As *information flows* can represent changes in *processes*, *entities*, or *systems*, we use these terms interchangeably in the rest of the discussion.

## Kinds of Scales

What kinds of scales are we talking about? The kinds of scales of interest are related to the kinds of observed variables (of interest to the observer). This, in turn, depends on the targeted domain of application or study, and on the appropriate scale for that application or study. For instance, the variables of interest to observation for physical objects are related to **space** and **time** (except for observations at the quantum scale – Cf. discussed below). The variables of interest for a computing system may be related to the usage of processing, storage and communication resources; to the number of component instances and their interrelations (i.e., structural models); or to the client-triggered call-paths through the system (i.e., behavioural models). These computing variables are all information-related, even if ultimately traceable to physical processes.

In physical studies, considering space and time as observable dimensions already implies a minimum abstraction, or granularity of observation, as these dimensions won't be found at the quantum scale [2]. Thus, space and time already represent abstract variables perceivable by some observers at "larger" scales. At smaller scales, we may have to consider instead the frequencies or energy needed for observation (out of scope). We can still consider space and time as the fundamental variables of observation for most physical systems of interest, unless we analyse quantum (computing) systems. Hence, it is useful to consider **spatial scales** and **temporal scales**.

Spatial scale typically represents a unit interval, area, or volume over which observations are made. Observations here are about spatial properties, structures, or shapes. Thus, spatial scale implicitly refers to *what* is being observed. Temporal scale refers to the frequency of observation. It refers to *when* observations are made, yet not to what is observed (e.g., spatial properties). At "higher" scales, information acquired via an observation can, in itself, represent variables of further observation, at even higher scales (i.e., granularity or abstraction). This is the case, for instance, when we monitor the information variables of computing processes (e.g., the state of a Java Object or of an e-commerce shopping basket). This means that we can talk about **information scales** (or abstraction levels).

**What does it mean for a system to operate at a certain scale?** We understand by this that the system's variables and their changes are observed at a certain granularity. The observers, which may or may not be part of the same system, may operate at the same scale, or at different scales. A system (or system-of-systems) where different parts operate at different scales is referred to as a **multi-scale system** – i.e. *a system observed at multiple granularities*. It implies the existence of some observers that perform some sort of **mapping**, or translation between scales: observing variables of entities that operate at a lower scale and providing an abstraction of these for entities that operate at a higher scale; or the opposite, from higher to lower scales. We refer to an entity operating a higher scale as a higher-scale entity (**macro**) and to an entity operating at a lower scale as a lower-scale entity (**micro**). Higher/lower and macro/micro are relative with respect to each pair of scales.

Generalising across physical and informational objects, and in line with the above definition of information, we reduce our focus to information scales only (and associated time scales of observation). Hence, multi-scale systems are ensembles of information flows that observe each other and that may be observed externally at different scales (micro and macro). As we saw above, observing information implies acquiring information about it. The scale of observation represents the interval, range, granularity or abstraction level at which information is acquired[4].

## How to measure abstraction?

Various approaches are possible for obtaining a higher level of abstraction, or granularity, at a higher scale, from lower abstractions or granularity at a lower scale. E.g., this can be achieved by **sampling** (at the size of one granule interval, like a unit area for space); by **aggregating** (over the area of each granule, for example by averaging); or by **modelling** (schematising the granular observation). This results in an abstraction of information from lower scales (micro) to higher scales (macro), where information details about the micro scale are lost at the macro scale. But how can we measure this information gap, or abstraction, between scales?

We propose using the notion of **information abstraction entropy** to quantify information loss between micro and macro scales. The abstraction entropy of a macro-entity's value is proportional to the number of micro-state combinations of micro-entities that could result in that macro-value (possibly weighted by the occurrence probabilities of these state combinations). E.g., consider a macro-variable that takes the average value of four integer micro-variables, which can take binary values. If that macro-value took the value of 0.5, then its abstraction entropy would be proportional to the six possible combinations of micro-values that could have produced that average: {0,0,1,1}, {1,1,0,0}, {0,1,0,1}, {1,0,1,0}, {1,0,0,1}, or {0,1,1,0}. Similarly, consider that the macro-entity samples the average value every two time steps versus every ten time steps. Then the less frequently sampled average would include micro-variable information across a larger interval of time and thus would have higher abstraction entropy compared to the case of more frequent sampling. Hence, the abstraction entropy measures *the amount of uncertainty about the micro scale that is hidden at the macro-scale*.

Another important aspect here is the **amount of information** that is acquired about the informational object (process, or entity). This is limited by the observation granularity, or scale. It can be considered as a further, intermediary abstraction between the observed object and the observer, imposed by the information flow that connects them. But how should we measure the amount of information extracted about an observed object? To be coherent with our definition of information as observable change, we could relate the amount of information as the amount of change being observed. For simplicity, we propose to reuse commonly employed information measurements related to the amount of resources needed to encode the information (i.e. logarithm in base two of the description length of an observed object), or to the amount of computing resources needed to store, process or communicate the information (i.e. in bytes, process cycles or bandwidth, respectively). This view is in line with our semantic definition of information, since we are still defining information flows as those producing a change in the observer (so our view is not purely material, but also functional, based on what that flow is doing).

---

[4] Acquired information typically contains: information about *what* is observed (e.g. which variables); information about *when* it was observed (e.g. time of observation); information about the state of what is observed at that time (e.g. values of variables).

## Multi-scale Feedback Systems

When information flows cycle between scales, they form **multi-scale feedback loops** [10]. Information about a lower scale (micro) is abstracted onto entities at a higher scale (macro), and this abstraction, in turn, is observed at the lower scale (micro), leading to adaptation. Hence, the higher scale observes the lower scale, and the lower scale observes the resulting abstraction from the higher scale (Figure 3). The observers of such multi-scale feedback systems are internal to the system (i.e. entities or processes that belong to the system).

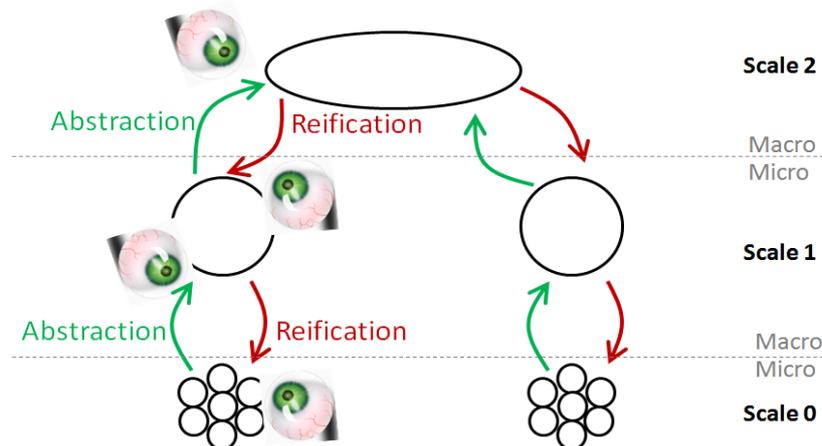

Figure 3: Multi-scale feedback system: observers are internal system entities or processes

The advantage of such multi-scale feedback loops lies in allowing individual micro-entities to access abstract information about the entire state of a set of micro-entities, possibly including themselves. Using abstract information instead of detailed information diminishes the amount of resources required to communicate, store and process it [11]. Therefore, it allows a large set of micro-entities to **coordinate** their actions, based on abstracted information about their collective state, and using a limited amount of resources (i.e. addressing H. Simon's "bounded rationality" problem [16]). The abstracted information can also be viewed as a means to reduce uncertainty for the micro-entities – as micro-entities can access this higher-level information, instead of having to extract detailed information at the micro-scale, they have more time and energy to spend on their own adaptation [15]. This allows for multi-scale systems to accommodate increasing levels of complexity (which doesn't mean that higher levels are more complex, but that the system as a whole can become more complex – able to handle more information flows and coordinate in the face of more changes).

As various inter-scale feedback loops operate in parallel, the relative **timing** of their operations becomes relevant [17]. In short, at a micro level, the larger the scope of abstract information received from higher-and-higher levels, the more delayed (and hence possibly outdated) this information may be. What matters here in terms of time is the relative order of events, or changes, observed at each local level, by each entity (i.e. which events occur before which other events); rather than the exact positioning of these events on a linear time line, which would have to be synchronised among all distributed entities involved.

Other important aspects of multi-scale systems include their topology, and the inter-scale abstraction and reification functions. These, in combination with various timing aspects, lead to different overall system properties, such as stability, convergence time, accuracy, maximum amplitude before reaching steady state, resilience.

## Summary and Future Perspectives

This short paper aimed to discuss and clarify the notions of *scale* and *multi-scale* as a conceptual basis for a theory of multi-scale systems in general and of multi-scale feedback systems in particular. It proposed an information-oriented approach to ensure generality and transferability across various application domains. We modelled systems as information flows, or change sequences, which merge and split to form higher or lower scales of information abstraction (macro) or detail (micro).

Future work will focus on refining and formalising concepts for quantifying inter-scale information abstraction (i.e. *abstraction entropy*, measuring the micro-macro gap, or hidden uncertainty) and information amount (i.e. resources needed to encode information at each scale). It will study how these general concepts can be adapted to various application domains. This includes exploring the optimal abstractions for each system considering its particular context and objectives. Inspiration will be drawn from analysing natural systems and gaining insights into how they answered the above questions. This will provide a stronger basis for a theory of multi-scale (feedback) systems, which aims to help scientists and engineers to understand, describe, analyse, develop and manage complex systems.